\documentclass[usenatbib]{mn2e}
\usepackage{times}
\usepackage{graphicx}

\title[Multicolour photometry of Balloon 090100001]
          {Multicolour photometry of Balloon 090100001: linking the two classes of pulsating hot subdwarfs}
\author[A.\,Baran, A.\,Pigulski, D.\,Kozie{\l}, W.\,Og{\l}oza, R.\,Silvotti and S.\,Zo{\l}a]
{A.\,Baran,$^{1,2}$\thanks{E-mail: andy@astro.as.wsp.krakow.pl (AB)}
A.\,Pigulski,$^{3}$ D.\,Kozie{\l},$^{4}$ W.\,Og{\l}oza,$^{1}$ R.\,Silvotti,$^{5}$ and S.\,Zo{\l}a$^{1,4}$\\
$^{1}$Mt. Suhora Observatory of the Pedagogical University, ul. Podchor\c{a}\.zych 2, 30-084 Cracow, Poland\\
$^{2}$Toru\'n Center for Astronomy, Nicholaus Copernicus University, ul. Gagarina 11, 87-100 Toru\'n, Poland\\
$^{3}$Instytut Astronomiczny Uniwersytetu Wroc{\l}awskiego, ul.~Kopernika 11, 51-622 Wroc{\l}aw, Poland\\
$^{4}$Astronomical Observatory of the Jagiellonian University, ul. Orla 171, 30-244 Cracow, Poland\\
$^{5}$INAF-Osservatorio Astronomico di Capodimonte, via Moiariello 16, I-80131, Napoli, Italy}
\begin{document}

\date{}

\pagerange{\pageref{firstpage}
\pageref{lastpage}} \pubyear{2005}

\maketitle

\label{firstpage}

\begin{abstract}
We present results of the multicolour $UBVR$ photometry of the high-amplitude EC\,14026-type star, Balloon\,090100001.   
The data span over a month and consist of more than a hundred hours of observations. Fourier analysis of these data
led us to the detection of at least 30 modes of pulsation of which 22 are independent.    The frequencies of 13 detected modes 
group in three narrow ranges, around 2.8, 3.8 and 4.7~mHz, where the radial fundamental mode, the first and second overtones
are likely to occur.   Surprisingly, we also detect 9 independent modes in the low-frequency domain, between 0.15 and 0.4~mHz.
These modes are typical for pulsations found in PG\,1716+426-type stars, discovered recently among cool B-type subdwarfs.
The modes found in these stars are attributed to the high-order $g$ modes.  As both kinds of pulsations are observed 
in Balloon\,090100001, it represents a link between the two classes of pulsating hot subdwarfs.  At present, it is probably 
the most suitable target for testing evolutionary scenarios and internal constitution models of these stars by 
means of asteroseismology.  

Three of the modes we discovered form an equidistant frequency triplet which can be explained by invoking rotational splitting of 
an $\ell$ = 1 mode. The splitting amounts to about 1.58~$\mu$Hz, leading to a rotation period of 7.1 $\pm$ 0.1 days.  

\end{abstract}


\section{Introduction}
Soon after the discovery of about a dozen pulsating B-type subdwarfs (sdB)
by the South African Astronomical Observatory astronomers [\citet{kilk97} and the next papers of the
series], two other groups (Bil\-l\'e\-res et al.~2002, {\O}stensen et al.~2001a,b, 
Silvotti et al.~2002a) undertook extensive searches for these 
interesting multiperiodic pulsators, called---after the prototype---the EC\,14026
stars.  The searches resulted in finding further variables, so that there are 32 EC\,14026 stars  
known at present.  Their pulsation periods range between 1.5 and 10 minutes, with a typical value of 2--3 minutes.
The amplitudes are rather small and rarely exceed 10~mmag.
Some extreme cases are known, however.  The longest periods 
are observed among the members having the lowest surface gravity: V\,338~Ser = PG\,1605+072 (Koen et al.~1998a,
Kilkenny et al.~1999), KL~UMa = Feige\,48 (Koen et al.~1998b, Reed et al.~2004a), HK~Cnc = PG\,0856+121 (Piccioni et al.~2000,
Ulla et al.~2001), V\,2214~Cyg = KPD\,1930+2752 (Bill\'eres et al.~2000), V\,391~Peg = HS\,2201+2610 ({\O}stensen et al.~2001a, 
Silvotti et al.~2002b), and HS\,0702+6043 (Dreizler et al.~2002).
Some of them (e.g., V\,338~Ser and HS\,0702+6043) have large amplitudes.
Recently, \citet{orei04} discovered another high-amplitude and cool EC14026 star, Balloon\,090100001 = 
GSC\,02248-01751 (hereafter Bal09).  The semi-amplitude of one of the detected modes amounts to 60~mmag
in white light, the largest amplitude known in an EC14026 star. With $B \approx$ 11.8~mag, 
the star is actually one of the brightest members of the group.

B-type subdwarfs are believed to be low-mass ($\sim$0.5~$M_\odot$) evolved stars that burn helium in their cores
(Heber 1986, Saffer et al.~1994).  
In the H-R diagram they populate the extended horizontal branch  (EHB) region.
As their hydrogen envelopes are very thin, they do not ascend the asymptotic giant branch 
and do not pass the planetary nebula 
phase of evolution \citep{grre90}. Instead, they evolve almost directly to the white-dwarf stage,
passing only through the hotter sdO phase \citep{dorm95}.   Their evolutionary past, however, 
is not well understood.  The discovery of pulsations in sdB stars opened therefore  a new way 
for investigating their internal structure by means of asteroseismology 
(see, e.g., Brassard et al.~2001).  
The effective temperatures of EC14026 stars range from 29\,000 to 36\,000~K, their surface 
gravities, $\log g$ = 5.2--6.1.  They do not occupy, however, a well-defined instability strip in the 
$\log T_{\rm eff}$--$\log g$ plane but are spread among non-pulsating sdB stars.   Curiously enough,
their pulsations were predicted almost at the time of discovery \citep{char96} as corresponding to radial and
nonradial low-order low-degree $p$ modes driven by the $\kappa$ mechanism.
The driving comes from the metal opacity bump that originates from the ionization of heavy metals, mainly
iron (Charpinet et al.~1997,  2001).

About two years ago, \citet{gree03} discovered low-amplitude multiperiodic light variations
with periods of the order of one hour among the coolest ($T_{\rm eff} <$ 30\,000~K) sdB stars.   
The periods of the members of this new class of variable stars
indicate that they are high-order $g$ modes.  Although \citet{gree03} announced the discovery 
of 20 stars of this type, detailed analysis was so far published only for the prototype, PG\,1716+426 
\citep{reed04a}.   The pulsations were explained by means of the same pulsation mechanism as the
short-period ones \citep{font03}, but with degrees $\ell$ = 3 or 4.    As far as the effective temperature and 
the character of excited modes are concerned, the two classes
of variable sdB stars resemble younger main-sequence pulsators, $\beta$~Cephei and SPB stars (see, e.g., Pamyatnykh 1999).
In the $\log T_{\rm eff}$--$\log g$ plane, the areas where the EC14026 stars and their long-period counterparts 
are found slightly overlap.  It seems therefore possible that both kinds of pulsations observed in sdB stars might coexist 
in a single star of this type.

The sdB nature of Bal09 was discovered by \citet{bixl91} using the balloon-borne SCAP ultraviolet
telescope. Its short-period variability was detected by \citet{orei04} by means of 
the 80-cm IAC80 telescope at Observatorio del Teide (Tenerife).
The effective temperature of 29\,500~K and $\log g \approx$ 5.3,
derived by the same authors, place it among the coolest and most evolved EC14026 stars.  From the data obtained on 
a single observational night, 
\citet{orei04} were able to detect only two independent modes.  Owing to the large amplitude and relatively long periods,
the star was an obvious target for follow-up time-series photometry.  We therefore decided 
to carry out multicolour observations of this star.  New observations were also obtained
by the discoverers (Oreiro, private communication).  In this paper, we present the results of our photometric run,
covering over a month.  We show that Bal09 exhibits both kinds of pulsations discovered in sdB stars.

\section{Observations and reductions}
We started observing Bal09 on June 30, 2004.  The bulk of the data was, however, obtained between August 17 and
September 19, 2004, on 18 nights.  All these observations were obtained at Mt.~Suhora Observatory 
with a 60-cm reflecting telescope, equipped with the ST-10XME CCD camera and Johnson-Cousins $UBVR$ filters.
The camera field of view was 6.8 $\times$ 4.5 arcmin and, in order to minimize storage time, the 3 $\times$ 3 pixels on-chip binning was
employed.  Except for the first night, June 30, when only $R$ filter was used, and the second night, 
when $V$-filter measurements were not made, 
the observations were carried out through four filters, $U$, $B$, $V$, and $R$.   The mean exposure times
amounted respectively to 15, 10, 7 and 7~s, resulting in about 1 minute-long filter cycle.  This enabled us to obtain roughly 6 points per main 
pulsation period of Bal09 in all four filters.  In total, 125 hours of observations were obtained in the $U$ and $B$ filters, 
119 in $V$, and 126 in $R$. 

\begin{figure}
\includegraphics[width=83mm]{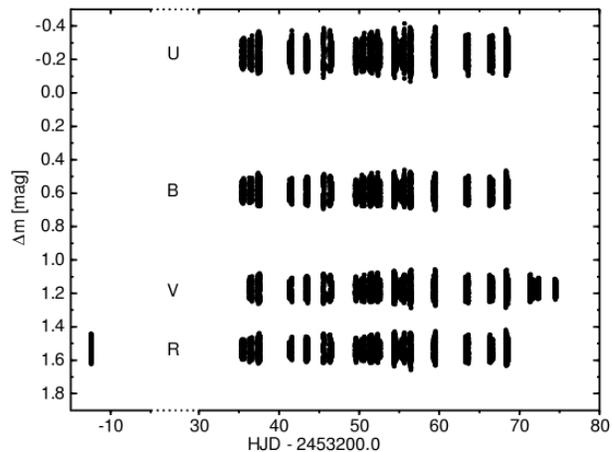}
\caption{Differential $UBVR$ magnitudes of Bal09 obtained during our run, plotted against
heliocentric Julian Day.  The observations were obtained at Mt.~Suhora Observatory, except those on the last 
three nights in $V$ that were taken at Loiano Observatory.}
\label{data}
\end{figure}

Additional observations were carried out on three observing nights between September 22 and 25, 2004, 
at Loiano Observatory with the 1.52-m Cassini telescope and the BFOSC CCD camera covering 13 $\times$ 12.6 arcmin on the sky.
In total, we gathered 13 hours of observations only in filter $V$.  The exposure times were 10 or 15~s long, depending on
sky conditions.  The time distribution of all observations is shown in Fig.~\ref{data}.

Over 7000 individual frames were obtained in $U$, $B$ and $R$ and about 8500 in $V$.
All these frames were corrected for bias, dark and flatfield in a standard way.  Then,  the magnitudes of
all stars detected in a frame were obtained by means of profile-fitting procedures of the
Daophot package \citep{stet87}.   The aperture magnitudes were also derived.  
In order to minimize photometric errors, the apertures were scaled with seeing.   
As a measure of seeing, we used the $\sigma$s of the two-dimensional Gaussian function that
described the analytical part of the point-spread function (PSF) calculated by Daophot.  The positions of
the stars were also taken from the PSF fits.  The seeing scaling factor, optimal from the point of view of the photometric errors, 
depended on the aperture magnitude of a star.   We derived the character of this dependence
from the post-fit residuals on a single good night and then applied it to all observations.  
As a comparison star, we chose the nearby, relatively bright star, GSC\,02248-00063.  It is brighter than Bal09 in $B$, $V$, and $R$, but
slightly fainter in $U$.   As can be seen in Fig.~\ref{data}, the mean differential magnitudes range from $-$0.23~mag in $U$ 
to 1.54~mag in $R$.  This large range reflects large differences in colour between Bal09 and the comparison
star.   Surprisingly, there is practically no difference between the mean differential magnitudes for the 
$V$-filter data from Mt.~Suhora and Loiano. This means that the instrumental systems in both sites are 
very similar and the $V$ data can be safely combined.

\begin{figure}
\includegraphics[width=83mm]{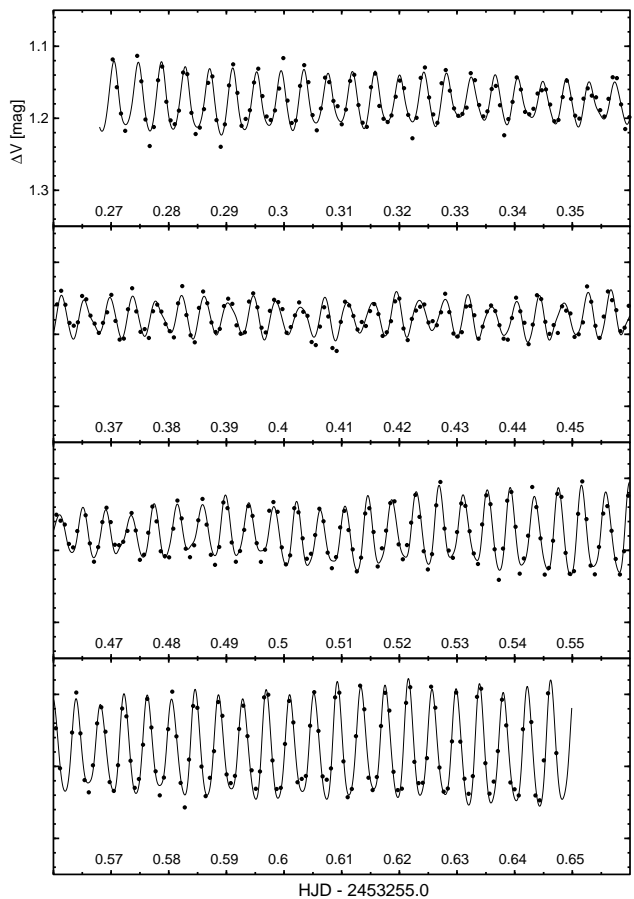}
\caption{Observations of Bal09 on a single night (September 6, 2004) obtained at Mt.~Suhora Observatory in filter $V$.  
The solid line is the synthetic light-curve consisting of 30 periodic terms listed in Table 1.}
\label{night55}
\end{figure}

The differential aperture photometry showed less scatter than the PSF photometry.
Consequently, for the frequency analysis that follows, we used the aperture photometry. 
One night's sample differential magnitudes in $V$ are shown in Fig.~\ref{night55}.

\section{Frequency analysis}
Prior to the frequency analysis, we rejected data points that had photometric
errors larger than a certain, arbitrarily chosen value, viz., 0.04~mag for $U$, 0.03~mag for $B$, and 
0.025~mag for $V$ and $R$.    As the amplitudes of the light variation of Bal09 are quite large, the next
three steps, i.e., rejecting outliers, correcting for second-order extinction effects and detrending, were done
using the residuals from a preliminary fit.  For this purpose, sequential prewhitening was done, until all periodic
terms with signal-to-noise (S/N) ratio\footnote{We define S/N as the ratio of the amplitude of a given peak
in the Fourier periodogram to the average amplitude in the whole range of the calculated frequencies. This definition
differs from the common definition of this value in the sense that usually the noise N is calculated from the periodogram of the 
very last residuals.  As we use S/N only for deciding when to stop prewhitening, this difference in defining S/N has no
influence on the final result.} 
larger than 6 were subtracted from the data.   

Having obtained the residuals, we first rejected all outliers
from the original data employing 5-$\sigma$ clipping.
Since the comparison star is much redder than Bal09, and we made observations through wide-band filters, second-order 
extinction effects can be expected.  In order to correct for them, we plotted the residuals, $\Delta m_{\rm res}$, 
obtained earlier, against the air mass, $X$, and derived the coefficients of the relation
$$\Delta m_{\rm res} = a(X-\mbox{1.25})+b$$ by means of the least-squares method.  The linear coefficient, $a$, was 
significantly different from zero and amounted to +0.044, +0.033, +0.009, and +0.012 $\pm$ 0.001 for $U$, $B$, $V$,
and $R$, respectively.   The $a(X-\mbox{1.25})$ factor 
was then subtracted from the original data.
Finally, the prewhitening was repeated and new residuals were computed.

Because 90\% of our data in $V$ and all data in the remaining three filters come from the same telescope/detector
combination, they constitute a very homogeneous dataset.  Nevertheless, some instrumental effects 
can produce artificial trends in the data, increasing amplitudes at the lowest frequencies, making it difficult
to say which low-frequency signals are real.  We will discuss this problem in Sect.~3.7 using data that
were not detrended.  At the present stage, detrending was performed, again using the residuals.  First, the average 
magnitudes in 0.07-day intervals were calculated and then interpolated using the cubic spline fit.  This
fit was subtracted from the original data. The 0.07-day 
interval was chosen by trial and error.  This choice secures that all signals at 
frequencies below $\sim$0.1~mHz are effectively reduced while those with higher frequencies, where $g$ modes are observed 
(Sect.~3.2), remain unaffected.

\subsection{Detected frequencies} 
\begin{figure}
\includegraphics[width=83mm]{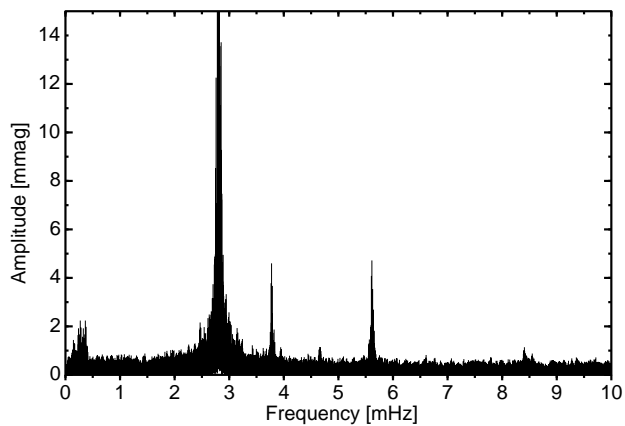}
\caption{Fourier amplitude periodogram of the detrended $V$-filter data of Bal09. 
The highest peak at about 2.8~mHz is truncated; its height is equal to about 53~mmag.}
\label{oriv}
\end{figure}

\begin{figure}
\includegraphics[width=83mm]{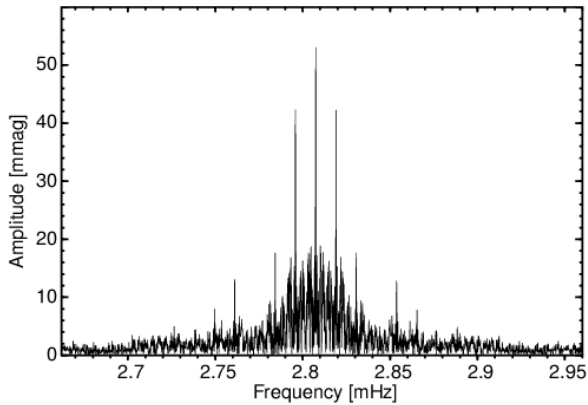}
\caption{Fourier amplitude periodogram of the artificial data containing a single sinusoidal 
term with frequency of 2.8074594~mHz and amplitude
of 53.3~mmag, distributed in time as the real observations in $V$.}
\label{swind}
\end{figure}

Detrended data were analyzed by means of the standard procedure that included calculating the Fourier
amplitude periodograms and sequential prewhitening of detected signals.  
At each step of prewhitening, the parameters of sinusoidal terms found earlier were improved by means of the non-linear 
least-squares method.  Data in each filter were analyzed independently.  The periodogram of the
$V$-filter data, up to the frequency of 10~mHz, is shown in Fig.~\ref{oriv}.   The structure of the aliases can be seen in Fig.~\ref{swind}, 
where we show Fourier periodogram of artificial data containing a single sinusoidal term with frequency and amplitude of the main mode
and distributed in time as the real observations in $V$. 
The daily aliases are quite high, as both sites where the $V$ observations were carried out are located at a similar longitude.
Therefore, the frequencies we derive, especially the low-amplitude ones, may suffer from the 1~(sidereal day)$^{-1}$ 
= 11.6~$\mu$Hz ambiguity.  Our data cover 35 days, resulting in a frequency resolution of about 0.5~$\mu$Hz \citep{lode78}.

The Fourier spectrum of Bal09 shown in Fig.~\ref{oriv} is dominated by the strong peak around 
$f \approx$ 2.8~mHz and its daily aliases. It is the main frequency discovered by \citet{orei04}.
Signals at the harmonics, 2$f$ (5.6~mHz) and 3$f$ (8.4~mHz), can also be seen.   The second frequency detected by these
authors, at about 3.8~mHz, is also well seen.  In addition, we see low-amplitude peaks at 4.7~mHz and, surprisingly, at low 
frequencies below 0.4~mHz.   Note that the peak at frequency of 0.3~mHz was already detected by \citet{orei04}, but these authors were
not able to attribute it convincingly to the star because of the very short time interval of their observations.
 
The analysis was carried out until all terms with S/N $>$ 4.5 were extracted.  The number of periodic terms found 
in this way was not the same in all filters, mainly due to the differences in the detection threshold.  The threshold was the lowest 
in $V$, where we found 27 independent terms and 7 combinations terms, including harmonics.   In $U$, we 
found 20 + 3, in $B$, 22 + 3, and in $R$, 23 + 6 terms.
Of all these terms, 17 (including 3 combination terms) were found in all four filters.  We finally adopted a model 
consisting of 22 independent terms and 8 combination terms.  The model comprised all terms detected in  
$V$ and at least one other filter, and all combination terms.  The frequencies of all terms included in this model and the results of the fit 
of this model to our multicolour time-series data are presented in Table \ref{fr30}.  
\begin{table*}
\centering
\caption{Frequencies, $UBVR$ semi-amplitudes, and phases of 30 sinusoidal terms adopted in the pulsation model of Bal09.  
Terms in a given frequency range are
listed according to decreasing $V$ amplitude.   Amplitudes in square brackets indicate that the term was not detected in the data 
obtained in a given filter.  Phases are given for the epoch HJD\,2453252.0, the errors of their last digits are given
in parentheses.  The entries at the bottom of the table 
give global parameters of the fit.  $N_{\rm obs}$ denotes the number of the data points
in a given filter, $\sigma_{\rm A}$, the rms error of the semi-amplitude, and $\Delta m$, the mean differential magnitude
(variable $-$ comparison).  Detection threshold corresponds to S/N = 4.5 in the 
Fourier periodogram of the residuals. }
\begin{tabular}{cclrrrrrrrr}
\hline\hline\noalign{\smallskip}
\multicolumn{2}{c}{Frequency}&\multicolumn{1}{c}{Period}&\multicolumn{4}{c}{Semi-amplitude [mmag]} &\multicolumn{4}{c}{Phase [rad]}\\
\multicolumn{2}{c}{[mHz]}&\multicolumn{1}{c}{[s]}&\multicolumn{1}{c}{$A_{\rm U}$} & 
\multicolumn{1}{c}{$A_{\rm B}$} & \multicolumn{1}{c}{$A_{\rm V}$} & 
\multicolumn{1}{c}{$A_{\rm R}$} &\multicolumn{1}{c}{$\phi_{\rm U}$} & 
\multicolumn{1}{c}{$\phi_{\rm B}$} & \multicolumn{1}{c}{$\phi_{\rm V}$} & 
\multicolumn{1}{c}{$\phi_{\rm R}$}\\
\noalign{\smallskip}\hline\noalign{\smallskip}
$f_{\rm A}$ & 0.272385 $\pm$ 0.000014& 3671.27 & 3.11 & 2.55 & 2.19 & 2.23 & 3.89(08) & 3.85(07)& 3.86(07)& 3.78(07) \\
$f_{\rm B}$ & 0.365808 $\pm$ 0.000012& 2733.67 & 3.29 & 2.81 & 2.13 & 2.30 & 2.80(08) & 2.87(06)& 2.74(07)& 2.76(07) \\
$f_{\rm C}$ & 0.325644 $\pm$ 0.000019& 3070.84 & 2.18 & 1.54 & 1.92 & 1.53 & 3.64(11) & 3.52(11)& 3.65(08)& 3.45(11) \\
$f_{\rm D}$ & 0.239959 $\pm$ 0.000020& 4167.38 & 2.24 & 2.00 & 1.59 & 1.87 & 4.84(12) & 4.85(09)& 4.79(10)& 5.01(09) \\
$f_{\rm E}$ & 0.15997 $\pm$ 0.00003& 6251.2 & 1.18 &[0.80]& 1.22 &[0.68] & 5.20(21) & 5.44(22)& 5.02(12)& 5.15(24) \\
$f_{\rm F}$ & 0.24629 $\pm$ 0.00004& 4060.3 &[1.08]&[0.67]& 1.14 & 0.86  & 4.79(25) & 4.44(28)& 4.80(14)& 4.15(20) \\
$f_{\rm G}$ & 0.29897 $\pm$ 0.00004& 3344.8 &[1.14]& 1.02 & 1.08 &[0.70] & 2.58(22) & 2.66(17)& 2.50(14)& 2.70(24) \\
$f_{\rm H}$ & 0.20194 $\pm$ 0.00003& 4952.7 & 1.42 & 0.98 & 1.04 &[0.48] & 0.26(18) & 0.54(18)& 0.22(15)& 0.50(35) \\
$f_{\rm I}$ & 0.22965 $\pm$ 0.00003& 4354.5 & 1.54 & 0.96 & 0.84 & 0.84  & 2.89(17) & 2.74(19)& 2.73(19)& 3.13(20) \\
\noalign{\smallskip}\hline\noalign{\smallskip}
$f_{1}$ & 2.8074594 $\pm$ 0.0000006 & 356.19393 &75.23 &57.71 &53.34 &50.26&2.099(04)& 2.112(03)&2.112(03)& 2.108(03)\\
$f_{2}$ & 2.8232252 $\pm$ 0.0000017 & 354.20483 &26.71 &21.92 &20.53 &19.92&1.137(10)& 1.149(09)&1.143(08)& 1.123(09)\\
$f_{3}$ & 2.824799 $\pm$ 0.000003 & 354.0075 &15.82 &12.86 &11.62 &11.57&5.359(16)& 5.347(14)&5.328(14)& 5.344(15)\\
$f_{4}$ & 2.826384 $\pm$ 0.000007 & 353.8090 & 5.86 & 5.12 & 4.71 & 4.59& 0.95(04) & 0.90(04)& 0.77(03)& 0.88(04) \\
$f_{5}$ & 2.85395 $\pm$ 0.00003 & 350.392 & 2.23 & 1.81 & 2.08 & 1.45 & 5.15(12) & 4.95(10)& 4.95(08)& 5.15(12)\\
$f_{6}$ & 2.85852 $\pm$ 0.00003 & 349.831 & 1.94 & 1.45 & 1.58 & 1.45 & 0.68(14) & 0.86(13)& 0.81(10)& 0.86(12)\\
$f_{7}$ & 2.85567 $\pm$ 0.00003 & 350.181 & 2.01 & 1.72 & 1.31 & 1.48 & 1.87(13) & 1.76(11)& 1.99(12)& 1.85(11)\\
\noalign{\smallskip}\hline\noalign{\smallskip}
$f_{8}$ & 3.776084 $\pm$ 0.000008 & 264.8246 & 5.94 & 4.73 & 4.16 & 4.07 & 1.16(04) & 1.15(04)& 1.16(04)& 1.13(04)\\
$f_{9}$ & 3.786788 $\pm$ 0.000017 & 264.0760 & 2.58 & 1.83 & 1.51 & 1.60 & 6.04(10) & 5.96(10)& 5.80(10)& 5.92(11)\\
$f_{10}$ & 3.79556 $\pm$ 0.00003 & 263.466 &[1.37]& 1.14 & 1.46 & 1.20 & 2.16(18) & 2.34(16)& 1.51(10)& 2.12(14)\\
\noalign{\smallskip}\hline\noalign{\smallskip}
$f_{11}$ & 4.64508 $\pm$ 0.00003 & 215.282 &[1.07]& 0.96 & 1.15 & 0.90 & 2.15(24) & 2.19(19) & 2.16(13)& 2.42(19)\\
$f_{12}$ & 4.66953 $\pm$ 0.00003 & 214.154 &[0.82]& 1.07 & 0.88 &[0.69]& 1.82(30) & 1.72(17) & 1.98(17)& 1.77(24)\\
$f_{13}$ & 4.66135 $\pm$ 0.00004 & 214.530 & 1.35 &[1.00]& 0.70 &[0.74]& 2.29(19) & 2.09(18) & 2.46(21)& 2.72(23)\\
\noalign{\smallskip}\hline\noalign{\smallskip}
2$f_{1}$ & 5.6149188 $\pm$ 0.0000012 & 178.09697 & 7.36 & 6.04 & 5.77 & 5.27      & 5.78(04) & 5.84(03) & 5.93(03)&5.88(03)\\
$f_{1}+f_{2}$ & 5.6306846 $\pm$ 0.0000018 & 177.59830 & 5.48 & 4.35 & 4.10 & 3.96 & 4.94(05) & 4.91(04) & 4.95(04)&4.92(04)\\
$f_{1}+f_{3}$ & 5.632259 $\pm$ 0.000004 & 177.5487 & 3.05 & 2.51 & 2.61 & 2.50    & 3.01(09) & 2.96(07) & 2.88(07)&2.96(07)\\
$f_{1}+f_{4}$ & 5.633843 $\pm$ 0.000007 & 177.4987 &[1.01]&[1.08]& 1.03 & 0.92    & 4.44(27) & 4.88(18) & 4.72(16)&4.53(20)\\
3$f_{1}$ & 8.4223782 $\pm$ 0.0000018 & 118.73131 &[0.85]&[0.68]& 0.95 & 0.96     & 3.53(30) & 3.59(26) & 3.87(16)&3.66(17)\\
$f_{1}-f_{\rm B}$ & 2.441651 $\pm$ 0.000012& 409.5589 &[1.06]&[0.75]& 0.82 &[0.73]& 2.79(23) & 2.79(24) & 2.55(18)&2.99(22)\\
$f_{2}+f_{3}$ & 5.648024  $\pm$ 0.000005  & 177.0531&[0.91]&[0.73]& 0.81 &[0.55]  & 2.37(29) & 1.93(26) & 2.43(20)&2.19(32)\\
2$f_{1}+f_{2}$ & 8.438144  $\pm$ 0.000003 & 118.5095&[0.69]&[0.42]&[0.58]& 0.94 & 2.58(37) & 2.10(43) & 3.11(27)&2.48(18)\\
\noalign{\smallskip}\hline\noalign{\smallskip}
\multicolumn{3}{r}{$\sigma_{\rm A}$ [mmag]} & 0.25    & 0.18   & 0.16   & 0.17  &&&&\\
\multicolumn{3}{r}{N$_{\rm obs}$}           & 7154    & 7197   & 8529   & 7382  &&&&\\
\multicolumn{3}{r}{Residual SD [mmag]}      & 14.80   & 10.54  & 9.59   & 9.96  &&&&\\
\multicolumn{3}{r}{Detection threshold [mmag]}   & 1.38    & 0.97   & 0.82   & 0.88  &&&&\\
\multicolumn{3}{r}{$\Delta m$ [mag]}        &$-$0.2290& 0.5940 & 1.1771 & 1.5389&&&&\\
\noalign{\smallskip}\hline
\end{tabular}
\label{fr30}
\end{table*}

\begin{figure}
\includegraphics[width=8cm]{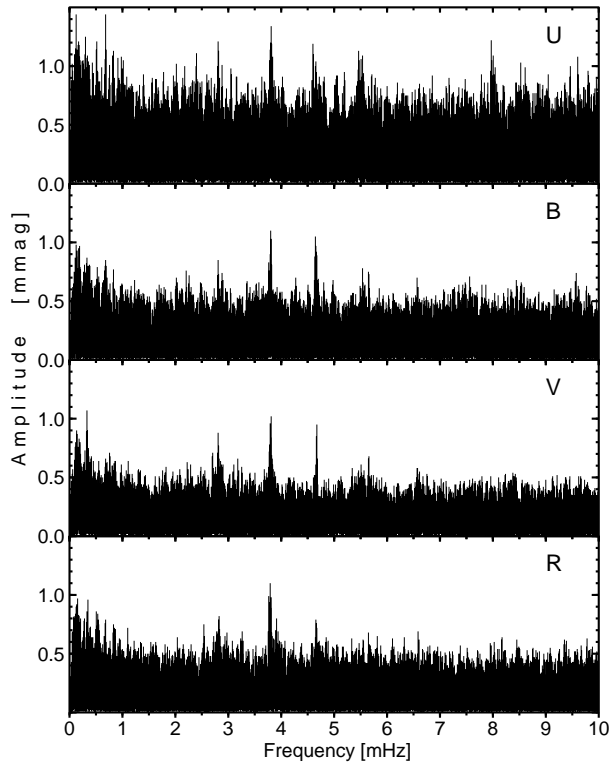}
\caption{Fourier spectra of the $UBVR$ residuals after removing 30 terms listed in Table 1.}
\label{res30}
\end{figure}

It is clearly seen from Table 1 (see also Fig.~\ref{data}) that the frequencies of modes detected in Bal09 cluster
in four narrow frequency ranges: below 0.4~mHz, around 2.8, 3.8, and 4.7~mHz.   After removing 
the 30 terms from the data, the Fourier spectra of the residuals still exhibit an excess power in these four 
ranges (Fig.~\ref{res30}).  Because the noise level increases towards low frequencies, we stopped extracting
terms with frequencies below 1~mHz.  However, extracting all terms down to S/N = 4.5 for frequencies in the range
1--10~mHz, led us to detecting additional 9 low-amplitude terms. Their frequencies are listed in Table \ref{fradd}. 
Note that only some are found in two bands, and even in these cases there is an ambiguity in frequency
because of aliasing.  It is clear that multi-site data are necessary to verify the reality of these terms.
\begin{table}
\centering
\caption{Additional high-frequency terms with S/N $>$ 4.5 found in the residuals.}
\begin{tabular}{cccccc}
\hline\hline\noalign{\smallskip}
 & \multicolumn{4}{c}{Detected frequency [mHz]} &\\
$f_i$& \multicolumn{1}{c}{$U$} & 
\multicolumn{1}{c}{$B$} & \multicolumn{1}{c}{$V$} & 
\multicolumn{1}{c}{$R$} & Comment\\
\noalign{\smallskip}\hline\noalign{\smallskip}
$f_{14}$ & --- & --- & 2.80599 & --- &\\
\noalign{\smallskip}\hline\noalign{\smallskip}
$f_{15}$ & --- & --- & --- & 3.76367 &\\
$f_{16}$ & --- & --- & --- & 3.79036 &\\
$f_{17}$ & --- & 3.80592 & --- & 3.79430 & daily aliases\\
$f_{18}$ & 3.80921 & 3.79767 & --- & --- & daily aliases\\
$f_{19}$ & --- & --- & 3.80509 & --- &\\
$f_{20}$ & --- & --- & 3.80670 & --- &\\
\noalign{\smallskip}\hline\noalign{\smallskip}
$f_{21}$ & --- & 4.64441 & --- & --- &\\
$f_{22}$ & --- & --- & 4.67491 & --- &\\
\noalign{\smallskip}\hline
\end{tabular}
\label{fradd}
\end{table}

We will now discuss frequency contents of the multicolour data of Bal09 in the four frequency
ranges mentioned above, illustrating the discussion with the Fourier amplitude spectra in the $V$ filter
obtained at different steps of prewhitening.

\subsection{Low-frequency region (0.15--0.4 mHz)}
\begin{figure}
\includegraphics[width=83mm]{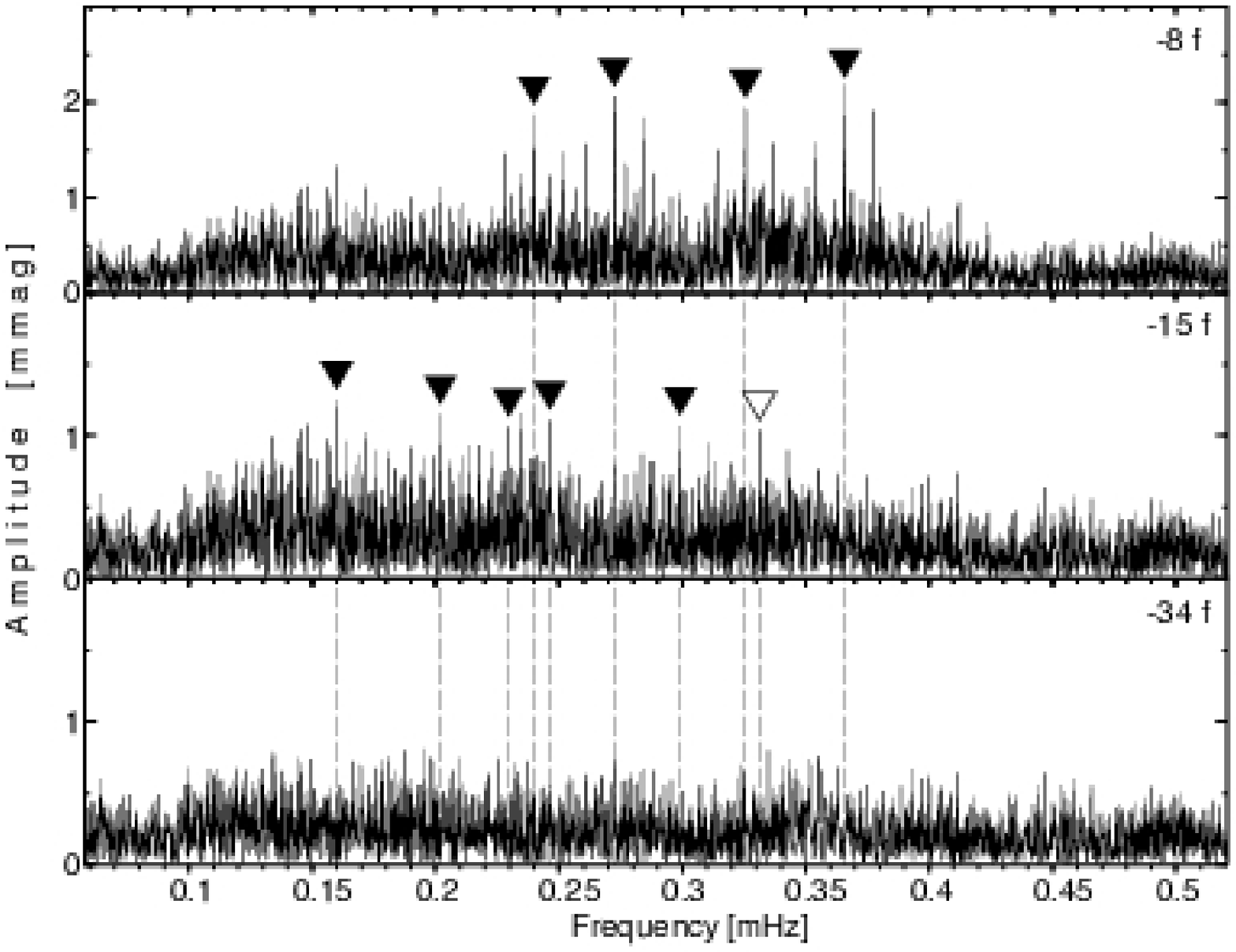}
\caption{Fourier amplitude spectra of the $V$ data of Bal09 in the low-frequency domain.  The number on
the right shows how many periodic terms were extracted from the data before calculating the 
periodogram.  Vertical dashed lines correspond to frequencies that have been already extracted.
The inverted filled triangles indicate frequencies included in the model given in Table 1, the open triangle, 
that detected only in $V$ data.}
\label{vpw1}
\end{figure}

Out of 22 independent terms included in the model listed in Table 1, nine were found in the low-frequency
range, between 0.15 and 0.4~mHz.  Five ($f_{\rm A}$ to $f_{\rm D}$ and $f_{\rm I}$) 
were detected in all four filters.   The Fourier spectrum in this range (Fig.~\ref{vpw1}) shows complicated structure; the first 
four terms have amplitudes of about 2~mmag in $V$, the remaining ones are even weaker.  
The dense spectrum of terms with low amplitudes in this range causes that aliasing is really a problem here.

\subsection{The 2.8~mHz region}
\begin{figure}
\includegraphics[width=83mm]{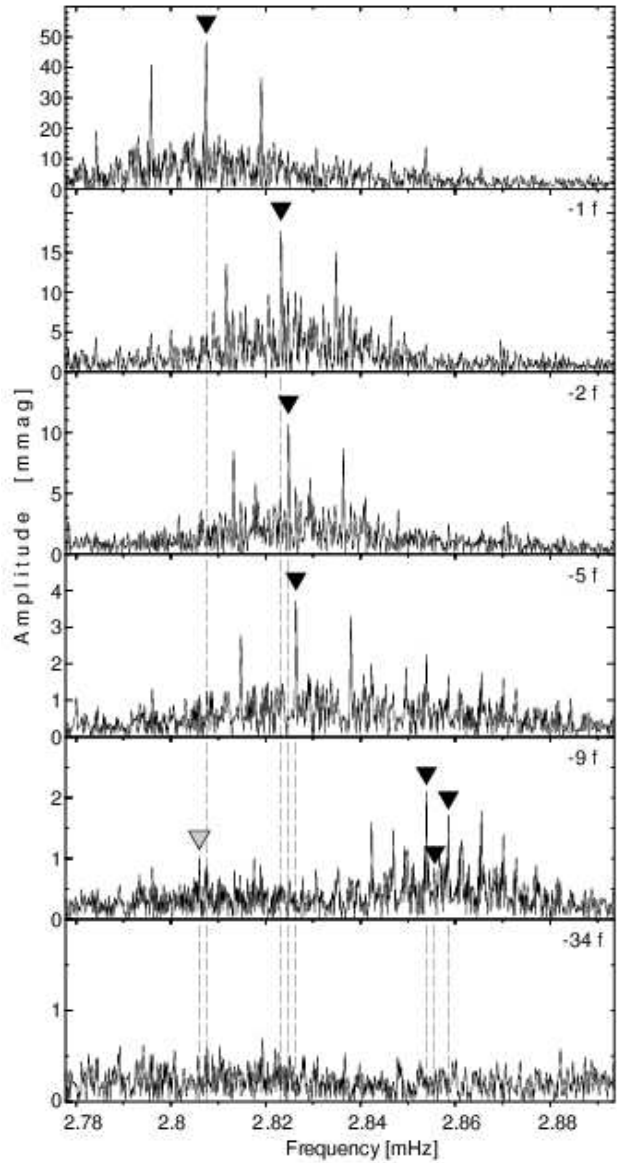}
\caption{Fourier amplitude spectra of the $V$ data of Bal09 in the vicinity of the main mode, $f_1$ = 2.80746~mHz.
Notation is the same as in Fig.~\ref{vpw1}, except that the frequency
listed in Table \ref{fradd} is indicated by the inverted gray triangle.  Note the differences in the ordinate scale.}
\label{vpw2}
\end{figure}

The strong peak at frequency of about 2.81~mHz, found by \citet{orei04}, is resolved in our data into several components.  
The strongest component has frequency $f_1$ = 2.8074594 $\pm$ 0.0000006~mHz (period 356.2~s) and is non-sinusoidal in shape,
as we detected two its harmonics.  All combination frequencies we detected, except $f_2+f_3$, involve $f_1$.  The 6-minute oscillations
corresponding to $f_1$ dominate the light curve of Bal09 (see Fig.~\ref{night55}).   We detected six terms in the vicinity of $f_1$
that were included in the 30-term model. 
One of the most interesting features seen in Fig.~\ref{vpw2} is that $f_2$, $f_3$, and $f_4$ form an equidistant frequency triplet.  
The frequency differences between the components of the triplet are the same within the errors and amount
to: $f_3 - f_2$ = 1.574 $\pm$ 0.003~$\mu$Hz and $f_4 - f_3$ = 1.585 $\pm$ 0.008~$\mu$Hz.  Thus, they can be
resolved only if observations cover more than a week.  It is natural to suggest that the triplet is a rotationally split one;
we will return to this interpretation in Sect.~5.3.  The beat period between $f_1$ and the 
triplet components is of the order of 0.7~d, and this beating is responsible for the changes of
amplitude in the observed light curve (Fig.~\ref{night55}).   

At a frequency about 0.03~mHz larger than that of the triplet, three next frequencies were
resolved. They are closely, although not equally spaced, $f_7 - f_5$ = 1.72 $\pm$ 0.04~$\mu$Hz, $f_6 - f_7$ = 2.85 $\pm$ 0.04~$\mu$Hz.

One additional term was found in the 2.8~mHz region in the residuals of the 30-term model 
(Table \ref{fradd}).  This $f_{14}$ term, found only in the $V$ data (Fig.~\ref{vpw2}), lies very close
to the combination frequency $f_1 + f_2 - f_3 \approx f_1 + f_3 - f_4$, 
but if real, it is more likely to be an independent mode.  The reason is that a  
much higher amplitude is found when solving for this frequency rather than assuming that it has a combination value.
The difference $f_1-f_{14}$ is equal to 1.46 $\pm$ 0.03~$\mu$Hz.

\subsection{The 3.8~mHz region}
\begin{figure}
\includegraphics[width=83mm]{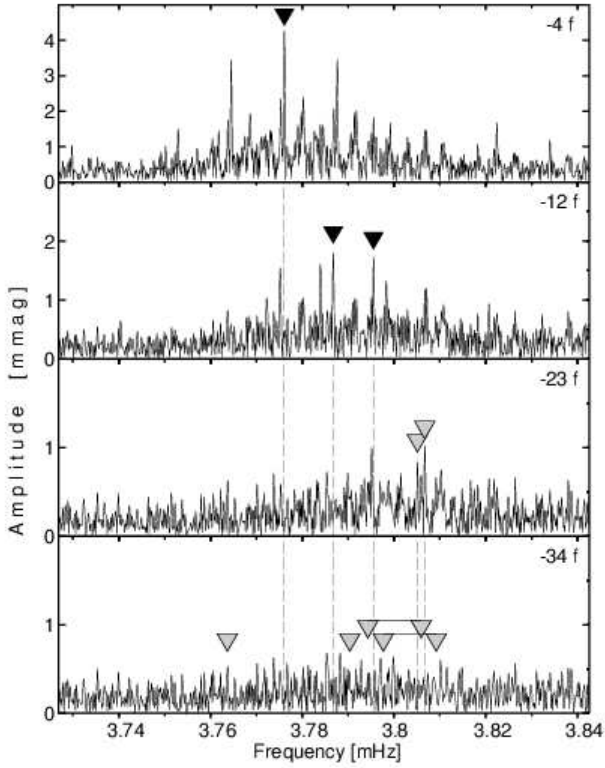}
\caption{Fourier amplitude spectra of the $V$ data of Bal09 around frequency 3.8~mHz, showing representative steps of prewhitening.
Symbols are the same as in Figs.~\ref{vpw1} and \ref{vpw2}. Note that 
out of six additional frequencies (Table \ref{fradd}), indicated by the gray triangles, two were found in the $V$ data only.
These two are shown in the second panel from the bottom, the remaining four, in the bottom panel. Triangles indicating alias frequencies 
are connected with horizontal lines.}
\label{vpw3}
\end{figure}

A peak at 3.78~mHz was already seen by \citet{orei04} in their Bal09 observations. We resolve it into three terms,
counting only those that appear in Table \ref{fr30} or nine, if those from Table \ref{fradd} are
taken into account.  It is interesting to note that $f_{20}-f_{19}$ = 1.61 $\pm$ 0.04~$\mu$Hz, a value that, to
within the errors, is the same as the separation of the three frequencies around 2.825~mHz.  As can be seen 
from Fig.~\ref{vpw3} (but see also Fig.~\ref{res30}), the region is rich in low-amplitude modes.  Not all can be
identified unambiguously from our data due to aliasing and the fact that many appear only in 
one periodogram.

\subsection{The 4.7~mHz region}
\begin{figure}
\includegraphics[width=83mm]{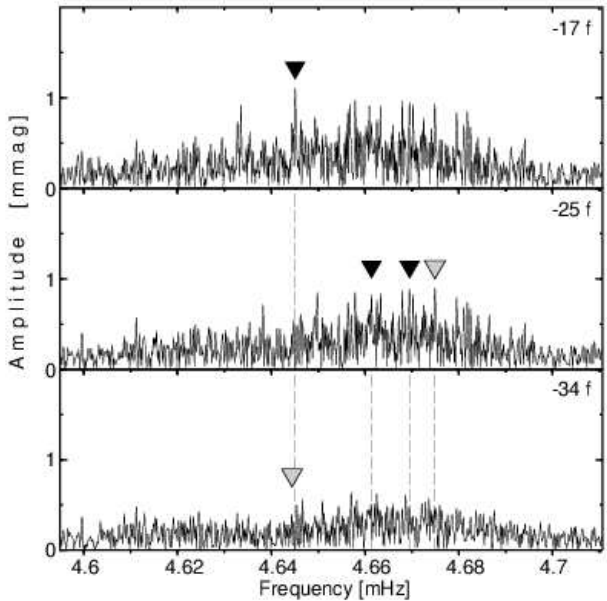}
\caption{The same as Fig.~\ref{vpw3}, but around frequency 4.7~mHz.}
\label{vpw4}
\end{figure}

Modes with frequencies around 4.7~mHz were not detected by \citet{orei04}.
We included three terms from this region in our 30-term model, but two more are detected with S/N $>$ 4.5: one in $B$ and one 
in $V$ (Fig.~\ref{vpw4}).  All terms in this region have very small amplitudes, around 1~mmag or even smaller.

\subsection{Combination frequencies}
\begin{figure}
\includegraphics[width=83mm]{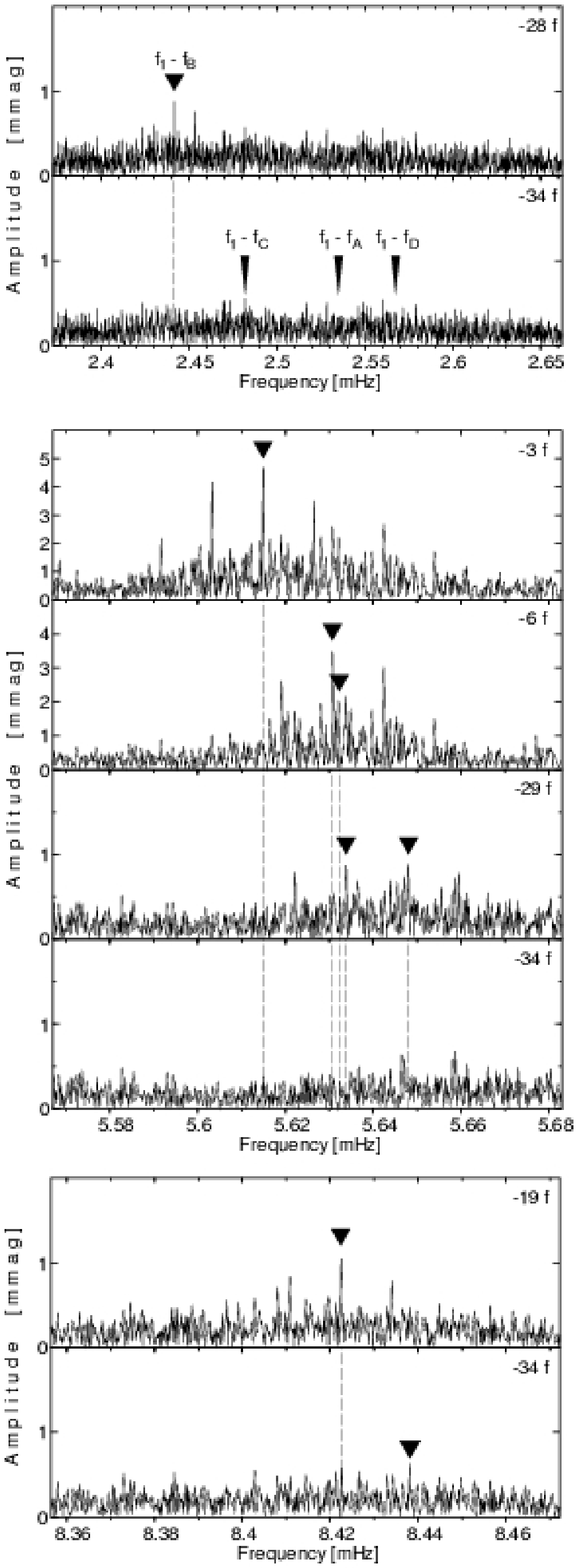}
\caption{The same as Fig.~\ref{vpw3}, but in three regions where the combination frequencies occur.  See text for
the explanation of the difference terms indicated in the top panel.}
\label{vpw5}
\end{figure}

The combination frequencies appear in three distinct regions (see Fig.~\ref{vpw5} and Table \ref{fr30}).  The
richest is the 5.6~mHz region where the 2$f_1$ harmonic and the sums of the highest-amplitude terms 
occur.  Five combination frequencies are detected in this region, including the sum of $f_1$ and all three 
triplet components, as well as $f_2 + f_3$.   

In the 8.4~mHz region,  the 3$f_1$ term is detected only in $V$.  However, although not detected  above S/N = 4.5, 
the 2$f_1$ + $f_2$ is also very well seen in the periodogram of the residuals (Fig.~\ref{vpw5}).

For the assessment of reality of the signals, the most important is, however, the detection of the 
difference term $f_1 - f_{\rm B}$ at 2.441651~mHz (see Fig.~\ref{vpw5}, top panel). This combination term involves
frequencies from the high and low-frequency ranges.  Its occurence proves 
that both $f_1$ and $f_{\rm B}$ originate in the same star.   In Fig.~\ref{vpw5} we have also indicated
the difference combination terms between $f_1$ and three other low-frequency terms with the largest amplitudes.
A detailed look at the residual spectrum in $V$ shows that in addition to the peak at $f_1-f_{\rm B}$ there is also a low 
but still noticeable peak at $f_1-f_{\rm C}$ = 2.48165~mHz.

\subsection{The lowest frequencies}
\begin{figure}
\includegraphics[width=83mm]{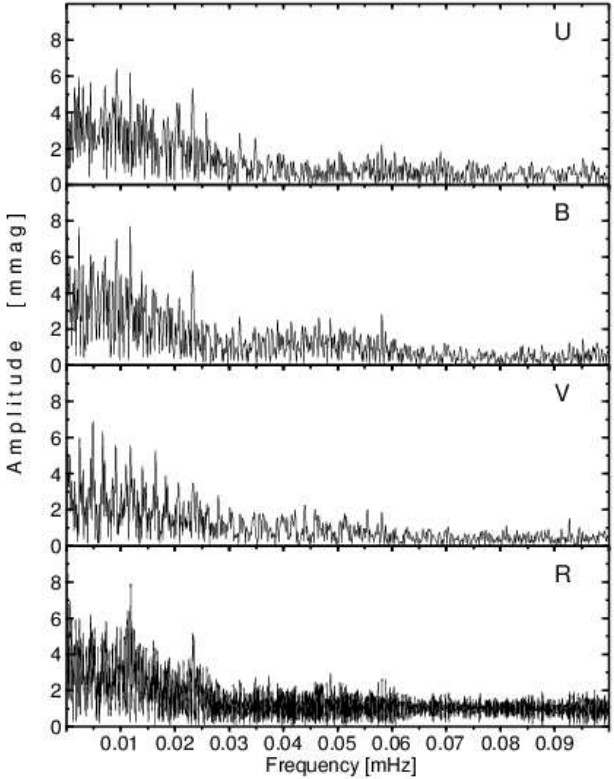}
\caption{The Fourier amplitude spectra of the Bal09 $UBVR$ observations for frequencies below 0.1~mHz.
The data were not detrended, but all terms listed in Tables \ref{fr30} and \ref{fradd} were 
removed.}
\label{lowfr}
\end{figure}

At the beginning of this section we mentioned that detrending was performed prior to the Fourier 
analysis.  This caused all information on signals with frequencies below $\sim$0.1~mHz to be
lost.  However, some real low-frequency signals may be present in the photometry of Bal09.
In particular, we can expect the difference combination frequencies.
Therefore, we decided to analyze also the data that were not detrended.  First, the already known 
terms with frequencies higher than 0.1~mHz were removed from the data, and the Fourier periodograms
of the residuals were calculated.  They are shown in Fig.~\ref{lowfr} for all four filters. 

We see that peaks close to 0 and 1~d$^{-1}$ = 11.6~$\mu$Hz appear in all periodograms indicating that some night-to-night
changes, presumably of instrumental origin, are present in the data.  This can be seen most clearly in the
$R$ data.   However, there is an indication of a significant peak at frequency of about 9.26~$\mu$Hz or its alias,
11.57 $-$ 9.26 = 2.31~$\mu$Hz in the $U$ and $B$ data.  The corresponding periods are equal to 1.25 and 5 days.  
It is difficult to say whether this is a 
real periodic change.   We think it is rather an artefact caused by transparency changes and/or instrumental effects.
The differential combination terms are not visible below 0.1~mHz; they are probably lost in the noise.

\subsection{Reality of the detected modes}
It is very important from the point of view of the results of this paper
to be certain that the modes we detected, especially the low-frequency ones, are real.  Moreover, we 
would like to be certain that they originate in Bal09, and not in the comparison star.  In order to verify the latter, we
analyzed differential photometry of the comparison star with respect to another star in the field, GSC\,02248-00425, called
thereafter the check star.  Unfortunately, this star is almost 4~mag fainter than the comparison star, so that only the $V$ and
$R$ photometry is usable for it.  The periodograms of the 
differential photometry of the comparison star with respect to the check star show no peak above S/N = 3.7 corresponding to 
the semi-amplitude of 3.0~mmag in $V$ data and no peaks above S/N = 4.1 corresponding to 1.9~mmag in the $R$ data.
This justifies that all modes with semi-amplitudes exceeding 2~mmag in $R$ or $3$~mmag in $V$ originate indeed in Bal09.
Among them, there are two low-frequency modes, $f_{\rm A}$ and $f_{\rm B}$. 
We cannot, however, exclude the possibility that some low-frequency modes originate in the comparison star.

An additional argument for the reality of the low-frequency modes may come from an independent detection 
of a given mode in the data split into two parts; see \citet{kilk99} for an example.  Consequently, 
we divided the photometry of Bal09 into two, roughly equal parts, before and after HJD\,2453254.  Then, the analysis of the two parts 
was performed independently.  This was done for data in all four filters.  Of 22 independent terms from Table \ref{fr30}, 
we detected 14 terms in both parts with S/N $>$ 4.5, or 16 if detection threshold was lowered to S/N = 4.
The remaining six terms, $f_{\rm F}$, $f_{\rm G}$,
$f_6$, $f_7$, $f_{11}$, and $f_{12}$, were detected in the periodogram of only one part of the data.  
This does not necessarily mean that these modes are spurious.  As a consequence of splitting the data, 
the noise level increases, masking the low-amplitude modes.

An amplitude change could be another reason for finding a mode in only one half of the data. In order to
check this, we compared the semi-amplitudes of the modes derived independently from the two parts of the data, but with
the frequencies fixed at the values given in Table \ref{fr30}.  The formal rms errors of
semi-amplitudes derived from the time-series data similar to our dataset are known to be underestimated by a factor of about two 
(Handler et al.~2000, Jerzykiewicz et al.~2005; see also Montgomery \& O'Donoghue 1999).  If we take this
into account, denoting the doubled formal rms errors of the semi-amplitudes in a giver filter $i$ by $\varepsilon_i$, we may
conclude that the differences in semi-amplitudes between the first and the second part of our data exceed 
3\,$\varepsilon_i$ only for $f_1$ in $U$ (3.2\,$\varepsilon_{\rm U}$), $V$ (4.7\,$\varepsilon_{\rm V}$), 
and $R$ (3.2\,$\varepsilon_{\rm R}$), and for $f_3$ in $U$ (4.0\,$\varepsilon_{\rm U}$) and $V$ (3.5\,$\varepsilon_{\rm V}$). 
For about 70\% of filter/mode combinations the differences were smaller than $\varepsilon_i$.  We may therefore 
conclude that there are no significant amplitude differences between the two parts of the data except for a marginal change
of the amplitude of two modes, $f_1$ (increase) and $f_3$ (decrease).

\section{The amplitudes}
A vast majority of the photometric observations of EC\,14026 stars was made without any filter.  
This was a compromise between the required photometric accuracy and the exposure time needed for the
very short periodicities observed in these variables.  However, the multicolour observations are highly desirable since they can 
support mode identification, a starting point for successful asteroseismology.  The best examples of the 
multicolour observations of EC\,14026 stars made so far are
the $UBVR$ data of \citet{koec98} for V\,2203~Cyg = KPD\,2109+4401, the $u'g'r'$ 
photometry of the same star and V\,429~And = HS\,0039+4302 = Balloon 84041013 \citep{jeff04},
four-band photometry of V\,338~Ser \citep{falt03}, and $UBVR$ observations of LM~Dra = PG\,1618+563B \citep{silv00}.

Since Bal09 is relatively bright and mode identification is one of the main goals of our study, 
we observed it through wide-band filters.  There is, however, a price.
The residual standard deviation of our $UBVR$ observations, a measure of typical accuracy of a single
data point, is quite large, and amounts to 0.015, 0.011, 0.010, and 0.010~mag, respectively (Table \ref{fr30}). 
Fortunately, owing to the large number of the observations we gathered, the detection threshold 
in Fourier periodograms is reasonably low, 0.8--1.4~mmag.  

\begin{figure}
\includegraphics[width=83mm]{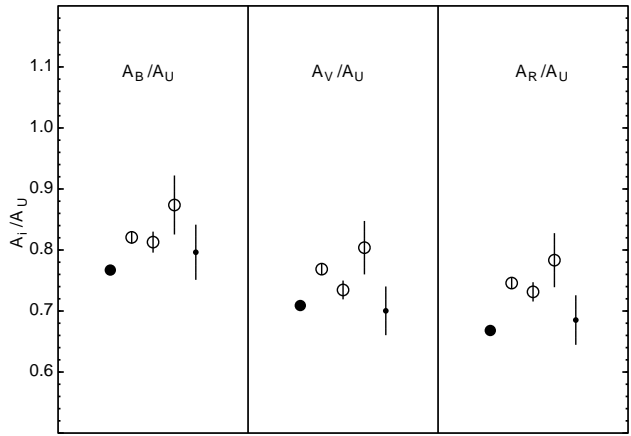}
\caption{Relative amplitudes, $A_i/A_{\rm U}$, $i = \{B,V,R\}$, for the five largest-amplitude modes detected in Bal09:
$f_1$, large filled circles, $f_2$, $f_3$, $f_4$ triplet, open circles, and $f_8$, dots.  Error bars 
were calculated from formal errors of semi-amplitudes given (Table \ref{fr30}).}
\label{amplr}
\end{figure}

The semi-amplitudes in all filters are given in Table \ref{fr30}.  It is worth noting that, along with the main mode in V\,338~Ser 
\citep{koen98a}, the $f_1$ mode has the
largest amplitude of all modes known in EC\,14026 stars. Generally, the semi-amplitudes become smaller towards 
longer wavelengths as the theory predicts (see, e.g., Ramachandran et al.~2004).  For 
five modes with the largest amplitudes we plot the
amplitude ratios in Fig.~\ref{amplr}. They are normalized to the $U$-filter amplitude.

The phase differences (Table \ref{fr30}) are equal to zero within the errors for all modes.

\section{Discussion}
\subsection{Preliminary mode identification}
There are at least two possibilities to identify modes solely from the photometry.  The first
method relies on the comparison of the amplitude ratios and phase differences with those calculated from 
the model atmospheres
for a given mode.  The amplitude ratios were already used to constrain degree $\ell$ in
EC\,14026 stars by \citet{koec98} and \citet{jeff04}.  However, these papers and the theoretical work
of \citet{rama04} indicate that when going from shorter to longer wavelength in the
optical domain, the amplitudes of low-degree ($\ell \leq$ 2) $p$ modes 
for EC\,14026 models behave similarly.  Therefore,
the $\ell$ for the $\ell\leq$ 2 modes is hardly derivable from amplitude ratios.  Modes with $\ell$ = 3 and 4 are 
much better separated: these with $\ell$ = 3 show rather flat dependence on wavelength, while those with $\ell$ = 4, much 
steeper than the $\ell\leq$ 2 modes.  The amplitude ratios for five modes shown in Fig.~\ref{amplr} exhibit similar
change with wavelength as the $\ell\leq$ 2 modes.  We may therefore rather safely interpret them as modes 
having $\ell$ equal to 0, 1 or 2.

The other possibility for mode identification from the photometry is the best match of the observed frequencies to those
calculated from the model [\citet{bras01} is a good example].  We 
make here only a rough comparison of the observed periods with those calculated in the currently 
available models, giving some indications as to the possible identifications.
A detailed mode identification including the stability analysis of models in a wide range of input parameters 
needs much better modelling and is beyond the 
scope of this paper; we postpone such a thorough analysis to a separate study.

\begin{figure}
\includegraphics[width=83mm]{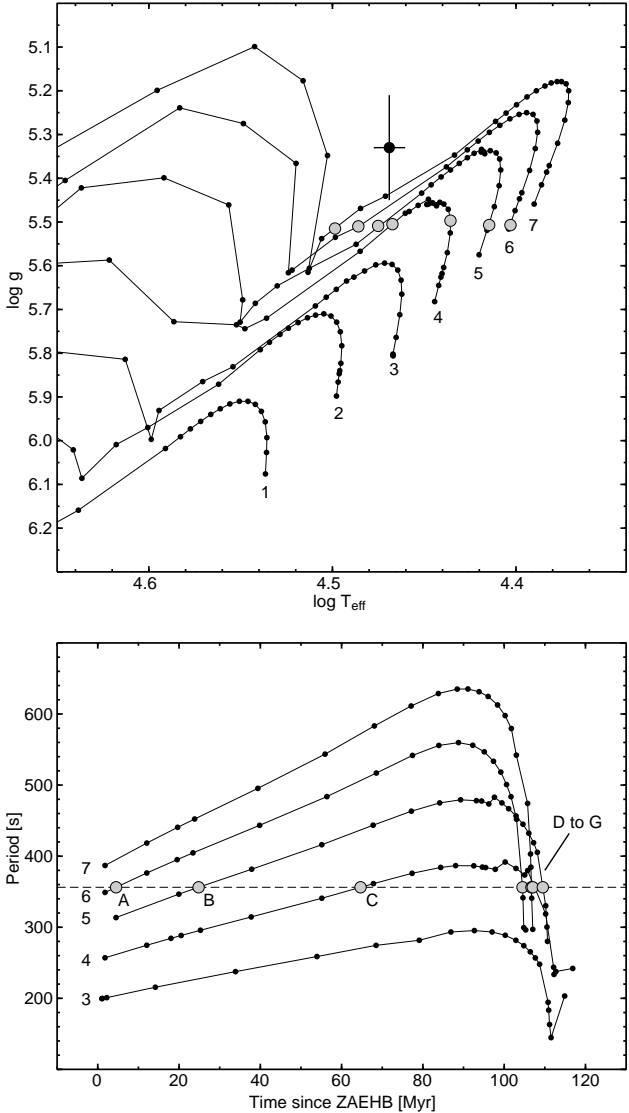}
\caption{{\bf Top}: Seven evolutionary sequences (labeled from 1 to 7) showing the extended horizontal branch evolution of
a 0.47--0.48~$M_\odot$ star, taken from \citet{char02}.  
It is mainly the hydrogen envelope mass that differentiate the models; it ranges
from 0.0001~$M_\odot$ for sequence 1 up to 0.0042~$M_\odot$ for sequence 7.  The large dot with error bars denotes the
parameters of Bal09 as derived by \citet{orei04}.  Seven filled gray circles show the position of models A to G for which
the period of the fundamental radial mode is equal to 1/$f_1$.\quad {\bf Bottom:} Periods of the fundamental radial mode, 
$\ell$ = 0, $p_0$ for five of the seven evolutionary sequences shown above.  The dashed horizontal line corresponts to $P_1$ = 1/$f_1$ =
356.19~s.  Gray filled circles show the positions of the seven models, A to G (see Table \ref{models}), that match the period
exactly.}
\label{permod}
\end{figure}

Let us start with the assumption that the main mode, $f_1$, is the radial fundamental one.  In addition to the 
expected wavelength-amplitude ration dependence, 
the justification for this assumption is its large amplitude. 
With this assumption, we can go to models in order to try to
interpret the other frequencies.  A large set of pulsational properties of evolutionary models was published 
by \citet{char02} (hereafter C02) for seven extended horizontal branch (EHB) sequences.  All sequences were calculated 
for a star with mass $\sim$0.48~$M_\odot$, but with different envelope masses: 0.0001, 0.0002, 0.0007, 0.0012, 0.0022,
0.0032, and 0.0042~$M_\odot$ for sequences 1 to 7, respectively. 
These models are shown in the $\log T_{\rm eff}$ -- $\log g$ diagram in Fig.~\ref{permod}.   The position of Bal09, according to
the parameters derived by \citet{orei04}, is also shown.

C02 provide also the pulsation periods for modes with $\ell$ up to 3. The periods of the fundamental radial
mode ($\ell$ = 0, $p_0$) are shown for all but the first two sequences in the bottom panel of Fig.~\ref{permod}.
The line for $P_1$ = 1/$f_1$ = 356.19~s intersects the model values in seven points corresponding to seven models 
belonging to sequences 4 to 7. 
Their properties, derived from linear interpolation of the model grid of C02, are listed in Table \ref{models}.
We see that none of the models with envelope mass less than $\sim$0.001~$M_\odot$ can be fit with 
the period of the fundamental radial mode equal to $P_1$.
The seven models in the sequence from the least evolved to the most evolved we denote with the letters A to G. 
The A to C models cover the main part of the EHB helium core burning evolution (HeCB in Table \ref{models}), 
D to F, its final stage, and 
G, the core helium exhaution (CHeE).  The models are plotted in Fig.~\ref{permod} with filled gray circles.  
Note that regardless of the model, the theoretical value of $\log g \approx$ 5.51 agrees reasonably well with the 
result of \citet{orei04}.  
\begin{table}
\centering
\caption{Parameters of interpolated models that have the period of the fundamental radial mode equal to 356.19~s.}
\begin{tabular}{ccrccc}
\hline\hline\noalign{\smallskip}
Model &C02 model & Age$^*$& Evol. &  $\log T_{\rm eff}$& $\log g$ \\
&sequence & [Myr]& phase &&\\
\noalign{\smallskip}\hline\noalign{\smallskip}
A&6 & 4.51 & HeCB & 4.403 & 5.51\\
B&5 & 24.81& HeCB & 4.415 & 5.51\\
C&4 & 64.66& HeCB & 4.436 & 5.50\\
D&4 &107.02& end of HeCB & 4.467 & 5.51\\
E&5 &109.51& end of HeCB & 4.475 & 5.51\\
F&6 &104.51& end of HeCB & 4.486 & 5.51\\
G&7 &106.70& CHeE & 4.499 & 5.52\\
\noalign{\smallskip}\hline
\end{tabular}
\label{models}

$^*$ since ZAEHB.
\end{table}

\begin{figure}
\includegraphics[width=83mm]{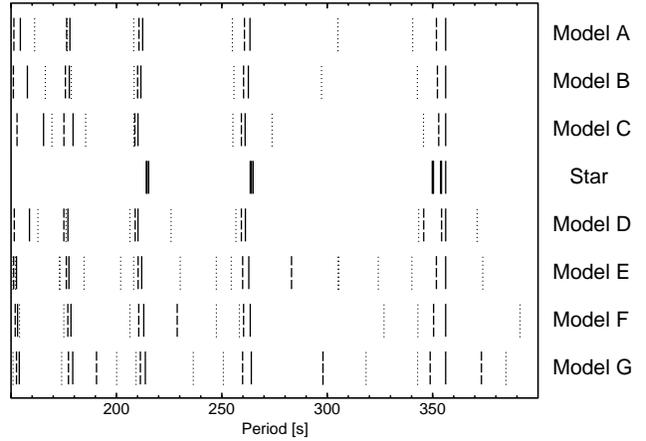}
\caption{Comparison of the pulsation periods for models A--G, indicated in Fig.~\ref{permod}, with periods detected in Bal09.
For the models, the radial ($\ell$ = 0) modes are shown with continuous lines, $\ell$ = 1 modes with dashed lines, 
and $\ell$ = 2 modes with dotted lines. See text for more details.}
\label{percomp}
\end{figure}

Having interpolated the models that match exactly $P_1$, we can try to compare the other periods as well.
This is done in Fig.~\ref{percomp} for all seven models.  For clarity, the $\ell$ = 3 modes are not shown in this figure.
We see that the observed pattern resembles to some extent the theoretical one (certainly with many 
high-overtone modes missing).  We can therefore derive some conservative conclusions based on this comparison.
Since besides the strong 356-s term, we have only three narrow ranges in period where the observed modes
occur, we can try to constrain the possible identification.

The separation between $f_1$ and the central term of the
rotationally split triplet ($f_3$) is best reproduced by model D.  This is also the model with no $\ell\leq$ 2 modes
having periods in the range between 270 and 340~s, in accordance
with observations. Moreover, model D is closest in $T_{\rm eff}$ to the observed 
value (Fig.~\ref{permod}).  However, we are far from indicating that model D is the best one, and constraining the 
evolutionary status of Bal09.  The problem requires better modelling.

Nevertheless, some indications as to the mode identification can be given even from this simple exercise. 
The mode closest in period to $p_0$, on the short-period side, is 
in all models the $\ell$ = 1 mode.  It has radial order $n$ = 1 for all models but G, where it has $n$ = 2.
However, while for the three less evolved models it is the $p_1$ mode, it is replaced in period for the remaining,
more evolved models, by $g_1$ (D--F) or even $g_2$ (G) due to the phenomenon of avoided crossing.
The $g$ modes under consideration have mixed character and are typical for evolved models also in 
main-sequence pulsators.  An $\ell$ = 1 identification can be therefore attributed with high probability to the
rotationally split modes $f_2$, $f_3$, and $f_4$; the azimuthal orders would then be $m$ = $-$1, 0, and +1, respectively.  

There is a general trend that for the three groups of periods detected in Bal09, that is, around 350, 260 and 215~s, there are always 
nearby theoretical modes with $\ell$ = 0, 1 and 2.  This can be seen in Fig.~\ref{permod}.  In Table \ref{mid}
we identified the theoretical modes closest to the observed ones for all seven models.
\begin{table}
\centering
\caption{Identification of the family ($p, f$ or $g$) and radial order (subscript) of 
modes with $\ell$ = 0, 1, and 2 closest to each of the three
period groups detected in Bal09 for seven models listed in Table \ref{models}.}
\begin{tabular}{cccccccccc}
\hline\hline\noalign{\smallskip}
& \multicolumn{3}{c}{Period group} & \multicolumn{3}{c}{Period group}& \multicolumn{3}{c}{Period group} \\
& \multicolumn{3}{c}{350--356~s} & \multicolumn{3}{c}{263--265~s} &\multicolumn{3}{c}{214--215~s} \\
Model & $\ell$ = 0 & 1 & 2 & 0 & 1 & 2 & 0 & 1 & 2 \\
\noalign{\smallskip}\hline\noalign{\smallskip}
A, B, C & $p_0$ & $p_1$ & $g_1$ & $p_1$ & $p_2$ & $p_1$ & $p_2$ & $p_3$ & $p_2$ \\
D & $p_0$ & $g_1$ & $g_1$ & $p_1$ & $p_2$ & $f$ & $p_2$ & $p_3$ & $p_2$ \\
E & $p_0$ & $g_1$ & $g_2$ & $p_1$ & $p_2$ & $f$ & $p_2$ & $p_3$ & $p_1$ \\
F & $p_0$ & $g_1$ & $g_3$ & $p_1$ & $p_1$ & $g_1$ & $p_2$ & $p_3$ & $p_1$ \\
G & $p_0$ & $g_2$ & $g_5$ & $p_1$ & $p_1$ & $g_2$ & $p_2$ & $p_2$ & $f$ \\
\noalign{\smallskip}\hline
\end{tabular}
\label{mid}
\end{table}

\subsection{Coexistence of low and high-frequency modes}
There is no doubt that the low-frequency modes, presumably high-order $g$ modes, exist in Bal09.
The question arises whether such modes are also present in the other cool and evolved EC\,14026 stars.  
The low-frequency region is sometimes ignored in the 
frequency analysis and signals, even when found, are frequently attributed to the atmospheric transparency
variations or removed during the detrending process.  However, high-order $g$ modes discovered so
far have periods shorter than 2 hours.  Thus, two to four cycles may be covered during a single night
and these signals can be easily distinguished from the transparency variations, especially in the 
differential photometry.  It is therefore worthwhile to reanalyze some already existing long data sets 
with the aim of searching for low-frequency modes.  While writing this paper, we have learned 
about the discovery of a long-period mode in HS\,0702+6043 \citep{schu04}.  
It is worth noting that both Bal09 and HS\,0702+6043 are neighbours of the PG\,1716+426 group of 
pulsating subdwarfs in the $\log T_{\rm eff}$ -- $\log g$ diagram, and some other EC\,14026 stars are close by  
(V\,338~Ser, V\,391~Peg and KL~UMa).  These stars should be considered 
as primary targets in the search for high-order $g$ modes in EC\,14026 stars.

\subsection{Frequency splitting}
Modes close in frequency were detected in many EC\,14026 stars, and some of them were supposed to be rotationally split
ones.  This would allow estimating rotational periods.  However, up to now, the only rotationally split triplet was found
in EO~Cet = PB\,8783 (O'Donoghue et al.~1998).  In several other EC\,14026 stars only close doublets were found.  
The equidistant triplet found by us in Bal09 is therefore the second one detected in an EC\,14026 star.  If we accept that it is a 
rotationally split $\ell$ = 1 mode (see Sect.~5.1), we can translate the observed mean splitting, $\Delta f$ = 1.58~$\mu$Hz, 
into the rotation period $P_{\rm rot} = (\mbox{1} - C_{n\ell})/\Delta f$,
where $C_{n\ell}$ is the Ledoux constant.  For all models considered in previous
section but model D, the values of $C_{n\ell}$ provided by C02 lie between 0.01 and 0.04,
implying $P_{\rm rot}$ = 7.1 $\pm$ 0.1~d.  Only for 
model D, the $C_{n\ell}$ is higher and can reach 0.2.  In that case $P_{\rm rot}$ would be smaller, about 6 d.
Bal09 is therefore very likely a slow rotator.
The 7.1-day rotation period for a star with a mass of 0.48~$M_\odot$ and $\log g$ = 5.51 results in the equatorial
rotation velocity of  1.44 km/s.

\subsection{Final remarks}
It is clear for us that Bal09 is one of the most interesting targets for
the future study by means of asteroseismology.   The main reason for this is the coexistence of the pulsations
inherent to both classes of pulsating sdB stars: the EC\,14026 and PG\,1716+426 stars. 
As far as the excited pulsations are concerned, it seems simpler than the best studied EC\,14026-type 
star V\,338~Ser that is a fast rotator 
\citep{hebe99}. Bal09 rotates slowly and is relatively bright.  
It is also exceptional because of the presence of a rotationally split mode
and the fact that a large number of combination frequencies are detected.

In addition to a better modelling we are going to undertake in near future, the next step in our understanding of Bal09 
could be taken if a new photometric and spectroscopic campaign were organized.  Such a campaign should
at least remove the 1 d$^{-1}$ ambiguity of some of the derived frequencies and lower the detection threshold
down to 0.1--0.2~mmag.  Both these goals could be achieved if the star were observed in several
observatories spread over the longitude, preferably in white light (or very wide band filters).  Multicolour observations
can support mode identification, but we showed in Sect.~4 that the $\ell$ value can be reliably constrained in that
way only for the largest-amplitude modes.  It seems, anyway, that even with accurate amplitudes it will
be hard to disentangle modes with $\ell$ = 0, 1, and 2, and it might be necessary to go to
satellite UV  to make a significant progress (see Fontaine \& Chayer 2004).

The campaign we plan should be supplemented by spectroscopic observations.  Some time-resolved spectroscopic observations of EC\,14026 
stars have been already done.  They were, however, usually too short in duration or badly distributed in time to result in a breakthrough in 
the mode identification.  In case of Bal09, however, even confirming that the main frequency, $f_1$, corresponds to the radial 
mode, as we suggested, would be helpful.   The best way towards proper identification would come, however, from accurate 
matching the model
frequencies to the observed ones.   This is even possible with our results provided that a thorough modelling would be done. 
As a by-product, stellar parameters of Bal09 could be obtained.  As noted above, we postpone this work to a separate paper.

{\it Acknowledgements.} The discussions with S.\,Ka\-wa\-ler, S.\,O'Toole and P.\,Moskalik and comments of an ano\-ny\-mous re\-fe\-ree 
are greatly appreciated.
We also thank M.\,Jerzykiewicz for critical reading of the manuscript and G.\,Kopacki for allowing us to use his 
calibration programs.

\bsp
\label{lastpage}
\end{document}